\documentclass[eqsecnum,aps,epsfig]{revtex4}
\usepackage{amsmath,amssymb,amsthm,bm}
\usepackage{psfrag}
\usepackage[dvips]{graphicx}
\usepackage[utf8]{inputenc} 
\usepackage{hyperref}

\begin{document}
\title{Contributions of the Cartan generators in potentials between static sources}
\author{Seyed Mohsen Hosseini Nejad}
\email{smhosseininejad@ut.ac.ir}
\affiliation{
Faculty of Physics, Semnan University, P.O. Box 35131-19111, Semnan, Iran}

\begin{abstract}

We investigate the contributions of the Cartan generators in the static potentials for various representations in the framework of the domain model of center vortices for SU($3$) gauge theory. Using the center domains with the cores corresponding to only one Cartan generator $H_8$, already given as a particular proposal, leads to some concavities in the potentials for higher representations. Furthermore, the string tension of the fundamental representation is the same at Casimir scaling and N-ality regimes. We add the contribution of the other Cartan generator $H_3$ to the potentials and therefore these shortcomings can be eliminated. However, we discuss that at intermediate range of distances the potentials induced by only  $H_8$ agree with the Casimir scaling better than those corresponding to both Cartan generators.    
\end{abstract}

\maketitle

\section{INTRODUCTION}\label{Sect0}
Non-perturbative regime of QCD is dominated by the phenomena of quark confinement and spontaneous chiral symmetry breaking. Lattice simulations \cite{DelDebbio:1996lih,Langfeld:1997jx,DelDebbio:1997ke,Langfeld:1998cz,Engelhardt:1999fd,Kovacs:1998xm} and infrared models \cite{Faber:1997rp,Greensite:2006sm,Engelhardt:1999wr,Engelhardt:2003wm,Deldar:2010hw,Deldar:2011fh,Deldar:2009aw,Nejad:2014hka,Nejad:2017hka} have shown that center vortices \cite{tHooft:1977nqb,Vinciarelli:1978kp,Yoneya:1978dt,Cornwall:1979hz,Mack:1978rq,Nielsen:1979xu}, quantized magnetic flux tubes in terms of the non-trivial center elements, cause the area law of the Wilson loop and therefore a linearly rising potential in the infrared regime. In addition, lattice simulations
have indicated that center vortices are also responsible for
the topological charge and spontaneous chiral symmetry breaking \cite{Faber:2017pfl,Faber:2020,deForcrand:1999our,Engelhardt:2002qs,Hollwieser:2008tq,Bowman:2010zr,Nejad:2016fcl,Hollwieser:2013xja,Nejad:2018pfl,Altarawneh:2020pfl}. Neglecting dynamical quarks in the vacuum, there is two non-perturbative ranges of interquark distances with distinct behaviors. At intermediate distances, the string tension follows the Casimir scaling rule \cite{Deldar:1999vi,Bali:2000un,Piccioni:2005un} while the asymptotic string tension only depends on $N$-ality \cite{Kratochvila:2003zj}. 

The domain structure model can account for
the color confinement in terms of the interaction of the Wilson loops with the center domains \cite{Faber:1997rp,Greensite:2006sm}. In the model, it is assumed that the QCD vacuum is filled with the domains of the center-vortex type corresponding to the non-trivial center element and of the vacuum type corresponding to the identity element. In some literatures given as a particular proposal for SU($3$) \cite{Deldar:2001dd} and SU($4$) \cite{Deldar:2005ss,Deldar:2007dr}, the core of domains is corresponding to only one Cartan generator which lead to some shortcomings. 

In this article, without this constraint and therefore using all Cartan generators within the core of domains, we study the static potentials. In the framework of the domain structure model, we investigate the Yang–Mills vacuum of the SU($3$) gauge theory including two Cartan generators $H_8$ and $H_3$. A convex potential without any concavity should be observed when interpolating between the short
distances and the asymptotic regimes \cite{Bachas:1985xs} while using only one Cartan generator $H_8$ for some higher representations unexpected concavity occurs \cite{Deldar:2001dd,HosseiniNejad:2019dd}. In addition, applying only one Cartan generator $H_8$, the string tension of the fundamental representation sources for SU($N$)($N=2,3,4$) is the same at intermediate and asymptotic distances \cite{Greensite:2006sm,Deldar:2001dd,HosseiniNejad:2019dd,Deldar:2005ss,Deldar:2007dr}. However there is no explanation for why the string tension of the fundamental representation should be the same in these regimes and it is mentioned as a weakness of the model in Ref. \cite{Greensite:2006sm}. This equality can not be explained by fine-tuning one of the parameters in the
model. We analyze the influence of increasing the number of Cartan generators within the core of domains on these shortcomings. Furthermore we investigate the contributions of the Cartan generators on the Casimir scaling of potentials at intermediate distances.

In Sec. \ref{Sect1}, we briefly explain the domain structure model. We analyze the contributions of Cartan generators in the potentials in Sec. \ref{Sect2}. The string tension of the fundamental representation sources is discussed in Subsec. \ref{Sect2-1}. Then, in Subsec. \ref{Sect2-2}, the Casimir scaling property is investigated. We discuss about the concavity of the potentials in Subsec. \ref{Sect2-3}. We summarize the main points of our study in Sec. \ref{Sect3}. Also, in appendix, the functions of the group factors for several representations which give some information about the details of potentials are presented.  

\section{Domain structures and quark potentials}\label{Sect1}
Any theory of confinement for SU($N$) gauge group should explain some features of the confining force which can be found in lattice simulations. The quark potential at intermediate distances, from the onset of the confinement to the onset of color screening, is expected to rise linearly.  In this regime, the string tension of the potential for the representation $r$ is proportional to $C_{r}$, the quadratic Casimir of the quark 
representation $r$, i.e. $\sigma_{r} \approx\frac{C_{r}}{C_{F}} \sigma_{F}$ where $F$ denotes the fundamental representation \cite{Deldar:1999vi,Bali:2000un,Piccioni:2005un}. At large distances, a pair of gluon-anti gluon is created in the vacuum and
combined with initial sources, and Casimir scaling breaks down and the potential depends on the $N$-ality $k_{r}$ of the representation i.e. $\sigma_{r} = \sigma(k_{r})$ where the string tension $\sigma(k_{r})$ is related to the lowest dimensional representation of SU($N$) with $N$-ality $k_{r}$. In addition, Lattice simulations \cite{Bachas:1985xs} have found that the static quark potential must be everywhere convex i.e. 
\begin{equation}
  \frac{d V}{d r}>0 \quad \text{and} \quad \frac{d^2 V}{d r^2} \leq 0.
\end{equation}
Therefore, a convex potential without any concavity should be observed when interpolating between the short
distances and the asymptotic regimes.

 Any theory of the quark confinement should be able to explain these features of the confining force. The domain structure model has been fairly successful in describing the mechanism of confinement in QCD \cite{Faber:1997rp,Greensite:2006sm}. The model assumes that the vacuum is filled with domains. In $SU(N)$, there are $N$ types of domains including $N-1$ types of center vortices corresponding to 
the non-trivial center elements of $z_n=\exp(i2\pi n/N)$ where the value $n=1,...,N-1$ and vacuum type corresponding to the trivial center element $z_0=1$ ($n=0$).
The effect of a domain on a planar Wilson loop is to multiply the loop by a group factor
\begin{equation}
\label{group factor}
{G}_r(\alpha^{n}_C(x))={1}/{d_r}\text{Tr}~\exp [i\vec{\alpha}^{n}_C\vec{{H}}],
\end{equation}
where $d_r$ is the dimension of the representation $r$, $\{H_i\}$ are generators of the Cartan subalgebra, and the angle $\alpha^{n}_C(x)$ depends on both the Wilson loop $C$ and the position of the domain $x$. If the domain is all contained within the Wilson loop 
\begin{equation}
\label{maximum1}
\exp [i\vec{\alpha}^{n}_C\vec{{H}}]=(z_n)^{k_{r}} \mathbb{I},
\end{equation}
 where $k_{r}$ is the $N$-ality of representation $r$. Using this constraint, one can obtain the maximum value of the angle ${\alpha}^{n}_{max}$. If the domain is outside the loop 
\begin{equation}
\label{maximum2}
\exp [i\vec{\alpha}^{n}_C\vec{{H}}]=\mathbb{I},
\end{equation}
and therefore the domain has no effect on the loop.

The quark potential induced by the domains is as follows \cite{Faber:1997rp,Greensite:2006sm}:
\begin{equation}
\label{potential}
V_r(R) = -\sum_{x}\ln( 1 - \sum^{N-1}_{n=0} f_{n}
[1 - {\mathrm {Re}}{G}_{r} (\vec{\alpha}^n_{C}(x))]),
\end{equation}
where $f_n$ is the probability that any given plaquette is pierced by an nth domain. An ansatz for the angle $\vec{\alpha}^{n}_C$ was introduced by Greensite $\it{et~ al.}$ \cite{Greensite:2006sm}. The color magnetic
flux within each
domain with square cross section $A_d=L_d \times L_d$ fluctuates almost
independently in small subregions of area $l^2\ll A_d$ which $l$ is a short length of correlation. The only restriction is that the total fluxes of the subregions 
must correspond to a center element of the gauge group. This square ansatz is as follows:

\begin{equation}
\label{Sansax}
         ({\alpha}^n_i(x))^2 = \frac{A_d}{ 2\mu} \left[
\frac{A}{ A_d} - \frac{A^2}{ A_d^2} \right]
                         + \left(\alpha^n_{i\{max\}} \frac{A}{ A_d}\right)^2,
\end{equation}
where $A$ is the cross section of the domain overlapping with the minimal area of the Wilson loop and $\mu$ is a free parameter. 

In the next section, we focus on the contribution of the Cartan generators to the static potentials between color sources for SU($3$)
gauge theory.

\section{Contributions of the Cartan generators in the static potentials}\label{Sect2}
The group factor given in Eq. (\ref {group factor}) plays an important role in the static potential and it is included $N-1$ Cartan generators $\{H_i\}$ in SU($N$) gauge group. For SU($3$) case, two Cartan generators $H_3$ and $H_8$ in the fundamental representation are as the following
\begin{equation}  
\label{h3}                                  
{H}_3=\frac{1}{2}\mathrm{diag}\big(1,-1,0\big),
\end{equation}
\begin{equation}
\label{h8} 
{H}_8=\frac{1}{2\sqrt3}\mathrm{diag}\big(1,1,-2\big).
\end{equation}
 
All diagonal elements of the Cartan generator $H_8$ are non-zero in the fundamental representation and the other Cartan generator $H_3$ comes from the $SU(2)$ gauge group. Because of this fact, as a particular proposal in Refs. \cite{Deldar:2001dd,HosseiniNejad:2019dd}, only one Cartan generator $H_8$ is used in the group factor and the Eq. (\ref {group factor}) is reduced as

\begin{equation}
\label{h8-group}
{G}_r(\alpha^{n}_C(x))= \frac{1}{d_{r}}
{\mathrm {Tr}} \exp[{\mathrm {i}}\alpha^n_{8}
H_{8}]. 
\end{equation}
Therefore, the core of a domain is related to only one Cartan generator $H_8$. In Refs.\cite{Deldar:2005ss,Deldar:2007dr}, for SU($4$) with three Cartan generators is also used only one Cartan generator with all diagonal elements non-zero in the group factor and the other two Cartan generators come from $SU(2)$ and $SU(3)$.

Now, we analyze the behavior of static potentials without the constraint that only one Cartan generator is used in the group factor. Therefore, in $SU(3)$, we include both $H_3$ and $H_8$ in the group factor as 

\begin{equation}
\label{h8-h3-group}
{G}_r(\alpha^{n}_C(x))= \frac{1}{d_{r}}
{\mathrm {Tr}} \exp[{\mathrm {i}} (\alpha^n_{3}
H_{3}+\alpha^n_{8}
H_{8})]. 
\end{equation} 

In the next subsections, we investigate the contributions of the Cartan generators on the properties of the confining force for SU($3$) gauge theory.
  
\subsection{The string tension of SU($3$) fundamental representation sources}\label{Sect2-1}

\begin{figure}[h!] 
\centering
\includegraphics[width=0.48\columnwidth]{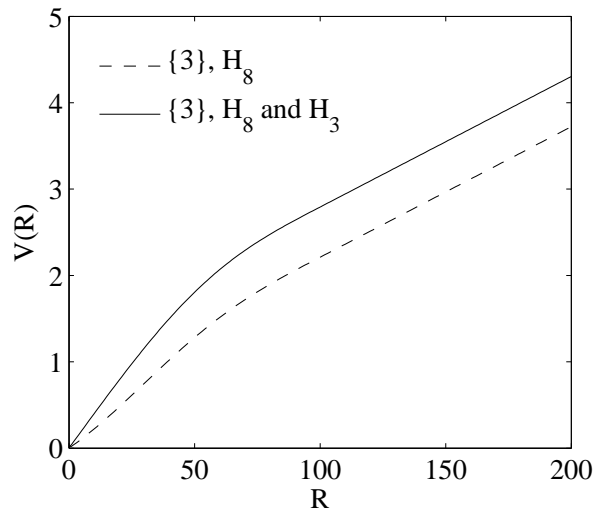}
\caption{ The static potentials induced by all domains in the fundamental representation. Using only the Cartan generator $H_8$, the string tension is almost the same at intermediate and asymptotic distances. Although $H_3$ contribution for large distances is zero, its contribution is non-zero at intermediate distances. Therefore, adding $H_3$ contribution to the potential, the string tension at intermediate distances is obtained different value with the one at asymptotic distances. The free parameters are $L_{d}=100$, $f_{1}=0.01, f_{0}=0.03$, and $L^{2}_{d}/(2\mu)=4$.   }
\label{FIG1}
\end{figure}

Applying the model in Refs. \cite{Greensite:2006sm,Deldar:2001dd,HosseiniNejad:2019dd,Deldar:2005ss,Deldar:2007dr}, using only one Cartan generator in the group factor of the potential, the string tension of the fundamental representation sources for SU($N$)($N=2,3,4$) is obtained the same at Casimir scaling and N-ality regimes. This equality can not be explained by fine-tuning one of the parameters in the
model \cite{Greensite:2006sm} and there is no explanation for why the string tension of the fundamental representation should be the same in these regimes. Now, we analyze the influence of increasing the number of Cartan generators within the core of domains on the potential of the fundamental representation. For SU($3$) gauge theory, we add the second Cartan generator in the potential. In SU($3$), there are two non-trivial
center elements in addition to the trivial center element
\begin{equation}
\mathbb Z(3)=\{z_0=1,z_1=e^{\frac{2\pi i}{3}},z_2=e^{\frac{4\pi i}{3}}\}.
\end{equation}
Since $z_1 = (z_2)^*$,  the vortex magnetic flux corresponding to $z_1$ is equivalent to an oppositely oriented vortex magnetic flux corresponding to $z_2$.
Using the Eq. (\ref {maximum1}) as well as the Cartan generators of $SU(3)$ given in Eq. (\ref {h3}) and Eq. (\ref {h8}), the maximum values of the angles $\alpha_{8\{max\}}^{(1)}$ and $\alpha_{3\{max\}}^{(1)}$ for the center vortices are equal to ${4\pi}/{\sqrt{3}}$ and zero respectively. Furthermore, these maximum values of the angles for the vacuum domains are zero. Therefore, the square ansatz given in Eq. (\ref {Sansax}) for the angles of the center vortices are:
\begin{equation}
\label{alpha-center}
         \Bigl(\alpha^1_8(x)\Bigr)^{2} = \frac{A_d}{ 2\mu} \left[
\frac{A}{ A_d} - \frac{A^2}{ A_d^2} \right]
                         + \left(\frac{4\pi}{\sqrt{3}}\frac{A}{ A_d}\right)^2,~~~
 \\
         \Bigl(\alpha^1_3(x)\Bigr)^{2} = \frac{A_d}{ 2\mu} \left[
\frac{A}{ A_d} - \frac{A^2}{ A^{2}_d} \right],                          
\end{equation}
and those for the vacuum domains are:
\begin{equation}
\label{alpha-vacuum}
         \Bigl(\alpha^0_8(x)\Bigr)^{2} = \frac{A_d}{ 2\mu} \left[
\frac{A}{ A_d} - \frac{A^2}{ A_d^2} \right],~~~
 \\
         \Bigl(\alpha^0_3(x)\Bigr)^{2} = \frac{A_d}{ 2\mu} \left[
\frac{A}{ A_d} - \frac{A^2}{ A^{2}_d} \right].                          
\end{equation}
Using the square ansatz, the static potential induced by all domains given in Eq. (\ref {potential}) for SU($3$) is as the following \cite{Greensite:2006sm}:     
\begin{equation}
         V_r(R) =  - \sum^{{ {L_d}/2 + R}}_{{x=-{L_d}/2}} \ln\Bigl\{(1-f_1-f_0) +f_1{\mathrm {Re}}G_{r}[\vec\alpha_C^1(x)+ f_0{\mathrm {Re}}G_{r}[\vec\alpha_C^0(x)]\Bigr\}, 
\label{Vr}
\end{equation}
where $f_1$ and $f_0$ are the probabilities that any given plaquette is pierced by a center vortex and a vacuum domain respectively. Figure \ref{FIG1} shows the static potentials $V(R)$ for the fundamental representation using all domains. The free parameters $L_{d}$, $f_{1}$, $f_{0}$, and $L^{2}_{d}/(2\mu)$ are chosen to be $100$, $0.01$, $0.03$, and $4$, respectively. We take the correlation length $l=1$ and therefore the static potentials are linear from the
beginning ($R=l$). As show, when we use only the Cartan generator $H_8$ in the group factor (Eq. (\ref {h8-group}), the string tension of the fundamental representation is almost the same at intermediate and asymptotic distances while by adding the Cartan generator $H_3$ within the group factor (Eq. (\ref {h8-h3-group}), the string tension of the fundamental representation at intermediate distances is different with the one at asymptotic distances. At intermediate distances, as shown in Eq. (\ref {alpha-center}) and Eq. (\ref {alpha-vacuum}), one can find the contribution of $H_3$ and the core of domains is corresponding to $H_3$ and $H_8$.
At large distances, the angle corresponding to $H_3$ is zero ($\alpha_3=0$) and therefore the string tension of N-ality regimes through adding $H_3$ contribution to the potential do not change. As a result, only Cartan generator $H_8$ contributes at large regimes while there are the contributions of both Cartan generators at intermediate regimes. Therefore, adding $H_3$ contribution to the potential, the string tension of the fundamental representation at intermediate distances can be achieved different value with the one at asymptotic distances, as shown in Fig.~\ref{FIG1}.

\subsection{Casimir scaling of SU($3$) static potentials}\label{Sect2-2}
Now, we study the contributions of Cartan generators on the Casimir scaling of potentials for some representations at intermediate distances. As mentioned above, although the potentials at large distances are governed by only $H_8$, one can add the contribution of $H_3$ at intermediate regimes. The group factors of the potentials within Eq. (\ref {Vr}) for the lowest representations of SU($3$) can be found in the appendix. Figure \ref{FIG2} plots the potential ratios $V_{\{6\}}(R)/V_{\{3\}}(R)$ and $V_{\{8\}}(R)/V_{\{3\}}(R)$ for different contributions of Cartan generators for center vortices at intermediate distances. The 
free parameters are $L_{d}=100$, $f_{1}=0.01$ and $L^{2}_{d}/(2\mu)=4$. Figure \ref{FIG3} is the same as Fig. \ref{FIG2} but for the vacuum domains where any given plaquette is pierced by a vacuum domain with the probability $f_0=0.01$. In the range $R\in [0,20]$, these potential ratios start out at the Casimir ratios ${C_{\{6\}}}/{C_{\{3\}}}=2.5$ and ${C_{\{8\}}}/{C_{\{3\}}}=2.25$.
\begin{figure}[h!] 
\centering
a)\includegraphics[width=0.48\columnwidth]{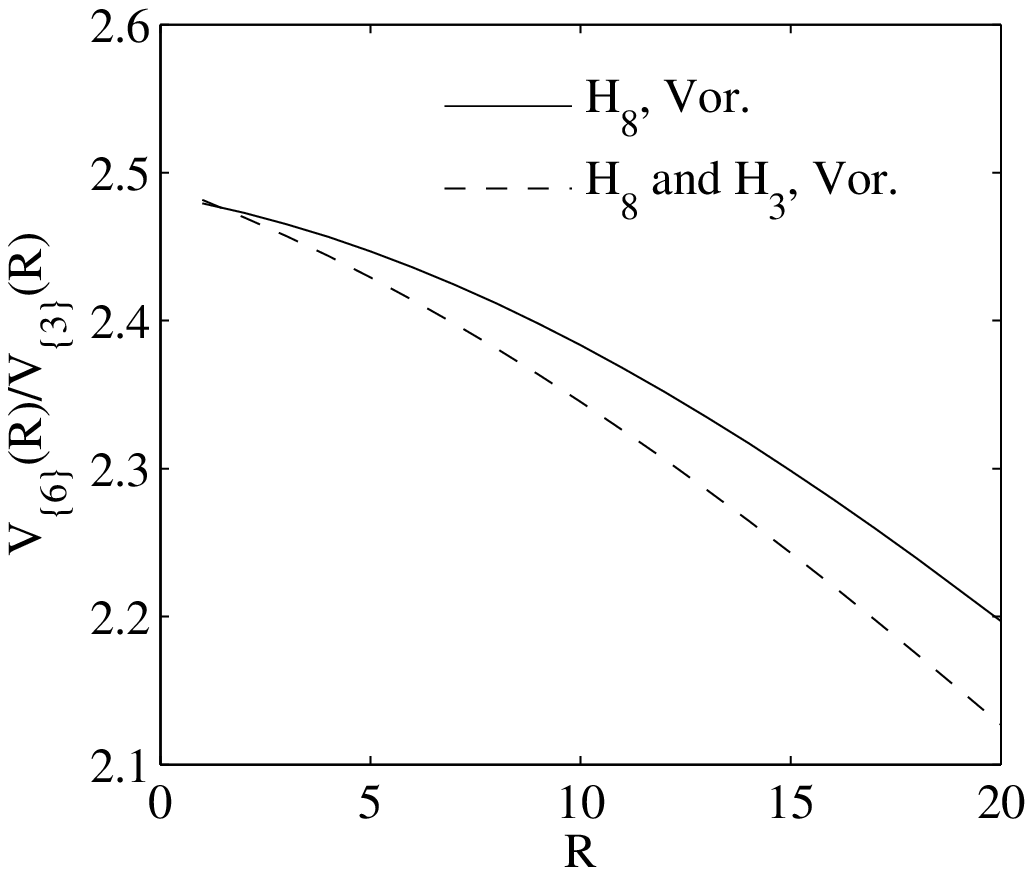}
b)\includegraphics[width=0.48\columnwidth]{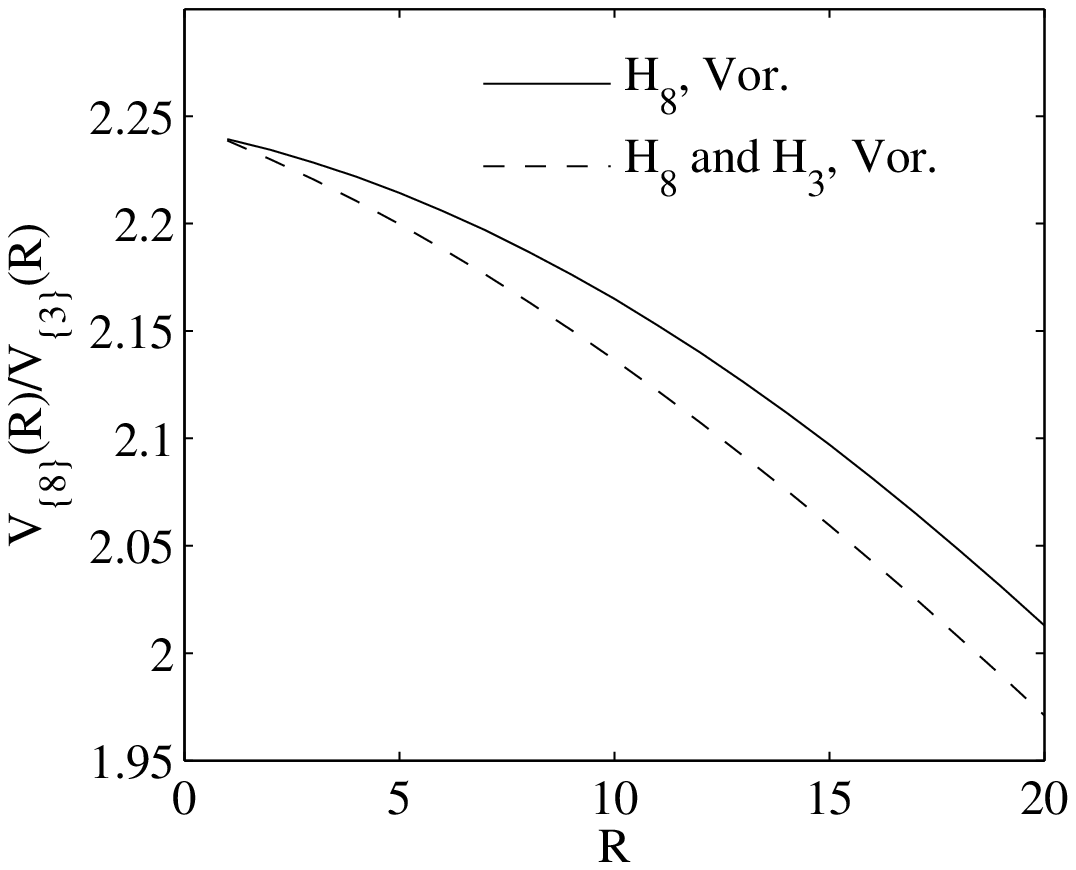}
\caption{ Potential ratios induced by center vortices for different contributions of the Cartan generators at intermediate distances a) for ${V_{\{6\}}(R)}/{V_{\{3\}}(R)}$ b) for ${V_{\{8\}}(R)}/{V_{\{3\}}(R)}$. In the diagrams, "Vor." means center vortices. Using only the Cartan generator $H_8$, the potential ratios which starts from the Casimir ratios violate slowly from the Casimir 
ratios. Adding the contribution of $H_3$ to the potentials, the slope of the ratios increases. The potentials induced by the center vortices corresponding to $H_8$ agree with the Casimir scaling better than those corresponding to both Cartan generators. The free parameters are $L_{d}=100$, $f_{1}=0.01$ and $L^{2}_{d}/(2\mu)=4$. }
\label{FIG2}
\end{figure}
\begin{figure}[h!] 
\centering
a)\includegraphics[width=0.48\columnwidth]{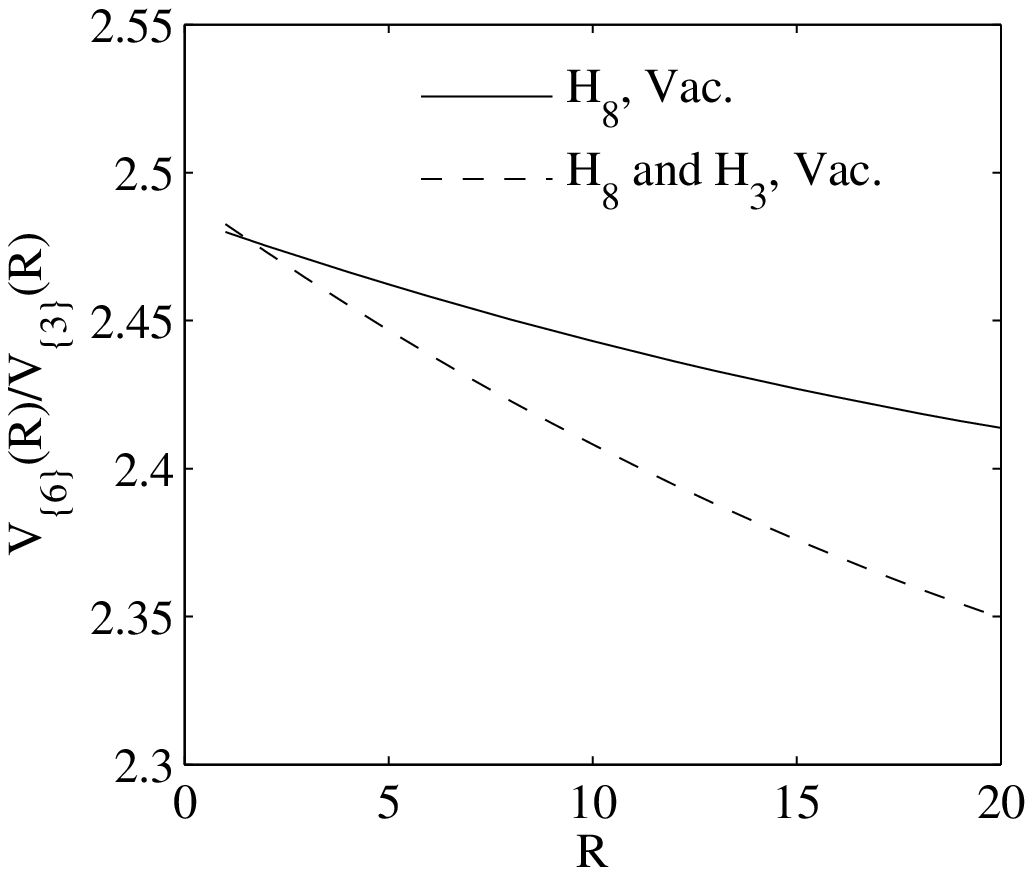}
b)\includegraphics[width=0.48\columnwidth]{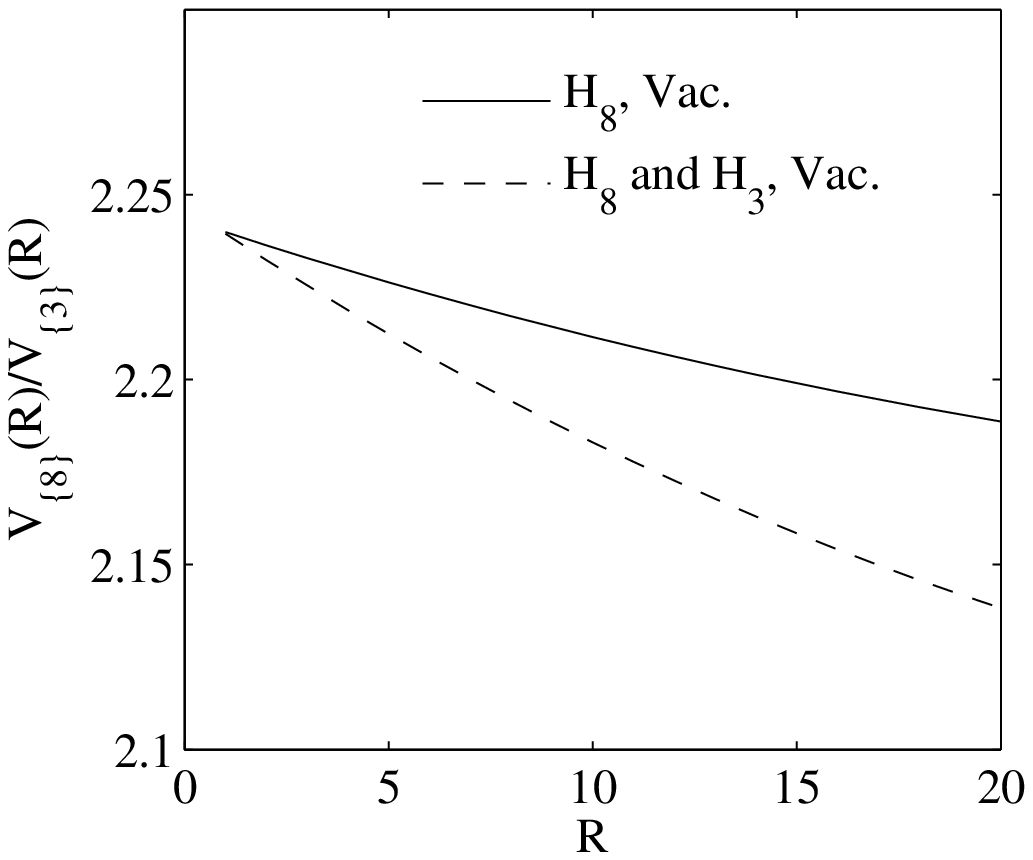}
\caption{ The same as Fig.~\ref{FIG2} but for the vacuum domains. In the diagrams, "Vac." means vacuum domains. The results are the same as those of center vortices. The free parameters are $L_{d}=100$, $f_{0}=0.01$ and $L^{2}_{d}/(2\mu)=4$. }
\label{FIG3}
\end{figure}
As shown in Fig. \ref{FIG2}, the potential ratios $V_{\{6\}}(R)/V_{\{3\}}(R)$ and  $V_{\{8\}}(R)/V_{\{3\}}(R)$ induced by the center vortices corresponding to only $H_8$ drop slowly from $2.5$ and $2.25$ to about $2.2$ and $2.02$, respectively. Adding 
the contribution of $H_3$ within the potential ratios $V_{\{6\}}(R)/V_{\{3\}}(R)$ and  $V_{\{8\}}(R)/V_{\{3\}}(R)$ obtained by the center vortices, the slope of the curve increases (the potential ratios change from $2.5$ and 
$2.25$ to about $2.12$ and $1.97$). In addition, as shown in Fig. \ref{FIG3}, for the same interval, the potential ratios $V_{\{6\}}(R)/V_{\{3\}}(R)$ and  $V_{\{8\}}(R)/V_{\{3\}}(R)$ induced by the vacuum domains corresponding to only $H_8$ drop very slowly from $2.5$ and $2.25$ to about $2.42$ and $2.18$, respectively. Adding 
the contribution of $H_3$ within the potential ratios $V_{\{6\}}(R)/V_{\{3\}}(R)$ and  $V_{\{8\}}(R)/V_{\{3\}}(R)$ obtained by the vacuum domains, the slope of the curve again increases (the potential ratios change from $2.5$ and 
$2.25$ to about $2.35$ and $2.14$). It should be noted that the same behavior occurs for the potential ratios for the higher representations.

As a result, the potentials induced by the domains with the core corresponding to $H_8$ agree with the Casimir scaling better than those obtained from the domains with the core corresponding to both $H_8$ and $H_3$. 

\begin{figure}[h!] 
\centering
\includegraphics[width=0.48\columnwidth]{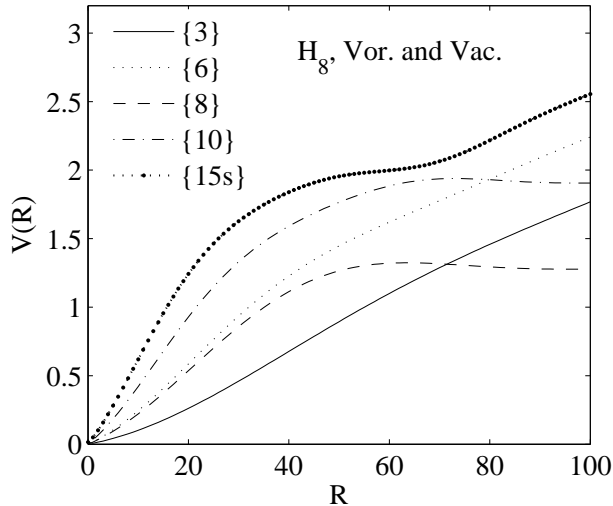}
\caption{ The static potentials induced by all domains corresponding to $H_8$ for several representations. Using only $H_8$ contribution in the potentials, the concavity is observed for the potential of the higher representation $\{15s\}$. The free parameters are $L_{d}=100$, $f_{1}=f_{0}=0.01$, and $L^{2}_{d}/(2\mu)=4$. }
\label{FIG33}
\end{figure}

Figure \ref{FIG33} plots the static potentials $V(R)$ induced by all domains corresponding to only Cartan generator $H_8$ for the lowest representations in the range $R\in [0,100]$. As shown, Casimir scaling and $N$-ality regimes for the representation $\{15s\}$ do not connect smoothly and some kind of unexpected concavity occurs which explicitly disagrees with
lattice results. In Ref. \cite{Deldar:2010hw}, using only center vortices corresponding to $H_8$, the concavity is observed independent of the ansatz for the angle in the representation $\{15s\}$.

In the next subsection, for removing the concavity of the representation $\{15s\}$, we study the effect of adding the contribution of $H_3$ on the potential.

\subsection{Concavity in the potential of the representation $\{15s\}$}\label{Sect2-3}

\begin{figure}[h!] 
\centering
\includegraphics[width=0.48\columnwidth]{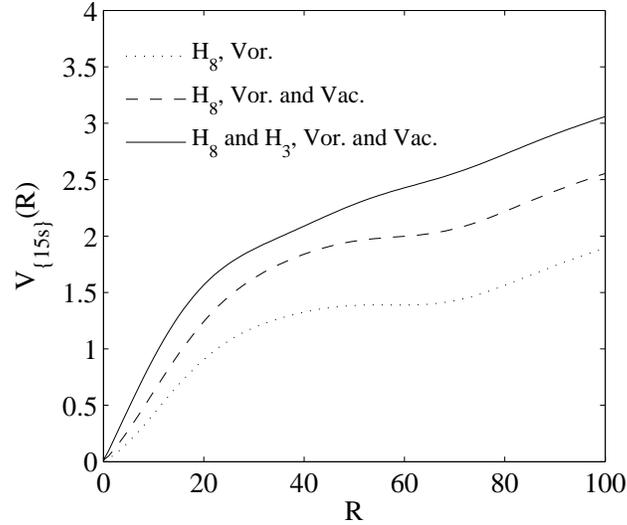}
\caption{ The potential of the representation $\{15s\}$ for various contributions of domains and Cartan generators. Using the center vortex contribution corresponding to $H_8$ on the vacuum, the concavity occurs for the potential. Adding the vacuum domain contribution corresponding to $H_8$, the concavity is reduced. However, this artifact stays on the potential. Using center vortices and vacuum domains corresponding to both $H_8$ and $H_3$, the concavity can be eliminated. The free parameters are $L_{d}=100$, $f_{1}=f_{0}=0.01$, and $L^{2}_{d}/(2\mu)=4$.  }
\label{fig:4}
\end{figure}

\begin{figure}[h!] 
\centering
\includegraphics[width=0.48\columnwidth]{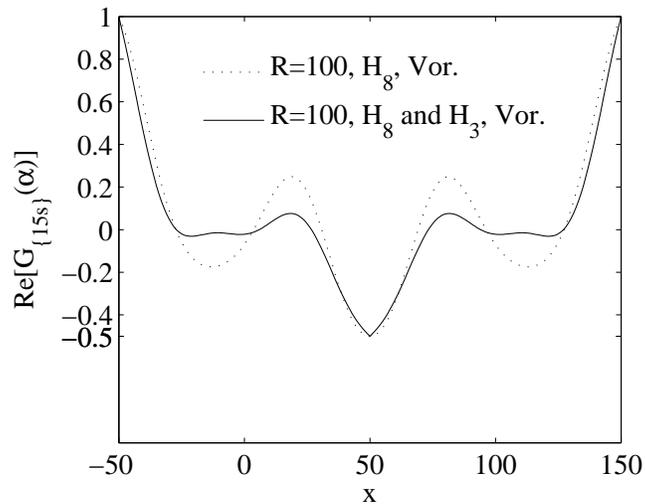}
\caption{The group factors ${\mathrm {Re}}G_{\{15s\}}(\alpha)$ of center vortices versus $x$ for the large size Wilson loop with $R=100$. The group factor, using only $H_8$ or both Cartan generators like the one of the fundamental representation, interpolates from $-0.5$, when the core of vortex is contained entirely within the Wilson loop, to $1$, when the core is entirely outside the loop. It shows that the string tension of N-ality regimes through adding $H_3$ contribution to the potential do not change. The free parameters are $L_{d}=100$ and $L^{2}_{d}/(2\mu)=4$. }
\label{fig:5}
\end{figure}

In the static potentials for all representations, the Casimir scaling and $N$-ality regimes should be connected smoothly to each other without any concavity. In Ref. \cite{HosseiniNejad:2019dd} using only one Cartan generator $H_8$, we add the contribution of the vacuum domain in the potential and therefore the concavity can be reduced. However this artifact stay and there is some concavities for higher representations. Now, we analyze the influence of adding the contribution of $H_3$ on the potential with the concavity especially for the representation $\{15s\}$. In Fig. \ref{fig:4}, the static potential for the representation $\{15s\}$ is plotted for various contributions of domains and Cartan generators.
As shown, the concavity of the potential induced by the domains with the core corresponding to $H_8$, can be removed through adding $H_3$ contribution in the potential. For the details of the potential in the representation $\{15s\}$, we analyze its group factors. In the appendix, the function of group factor of the representation $\{15s\}$ using both Cartan generators is given. Figure \ref{fig:5} plots the group 
factors of the representation $\{15s\}$ induced by center vortices for different Cartan generator contributions at large distances ($R=100$). The Wilson loop legs are located at $x=0$ and $x=100$. The group factor using only $H_8$ or both Cartan generators like the one of the fundamental representation, interpolates from $-0.5$, when the vortex core is located entirely within the Wilson loop, to $1$, when the core is entirely outside the loop. Also, in the same interval, when the core of the vacuum domain is located entirely within the loop, the group 
factor reaches to the trivial value $1$.
\begin{figure}[h!] 
\centering
a)\includegraphics[width=0.48\columnwidth]{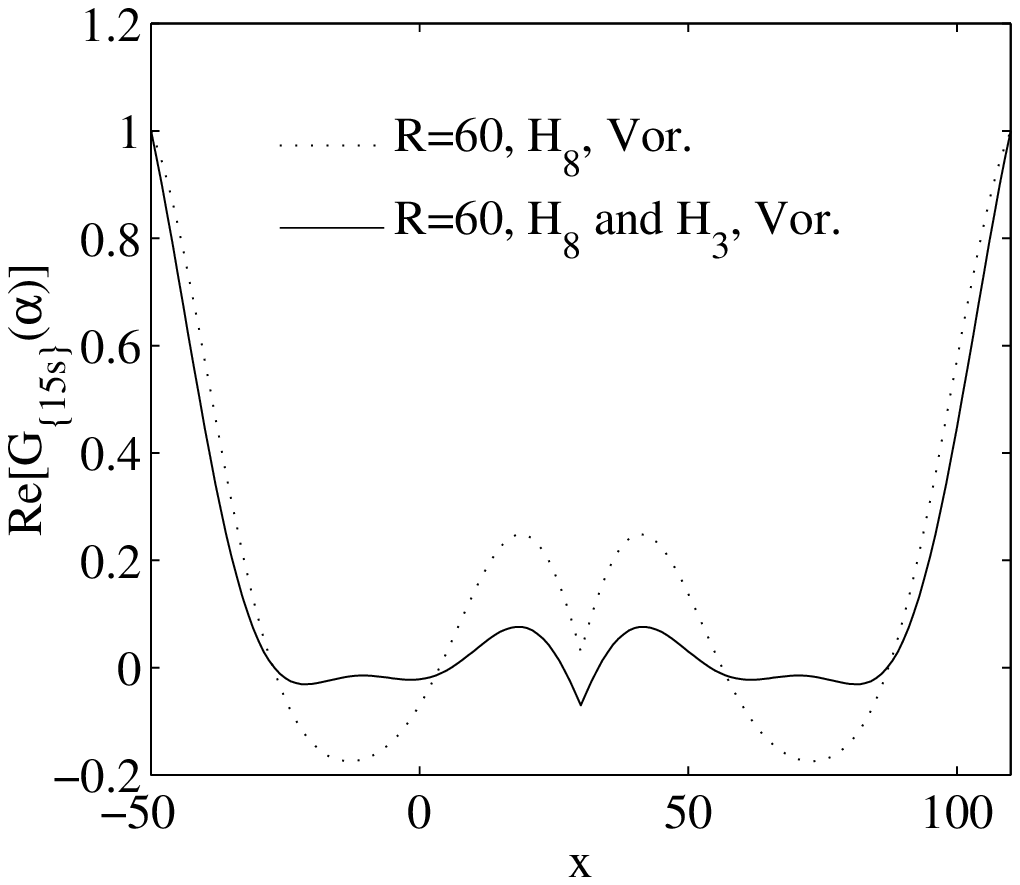}
b)\includegraphics[width=0.48\columnwidth]{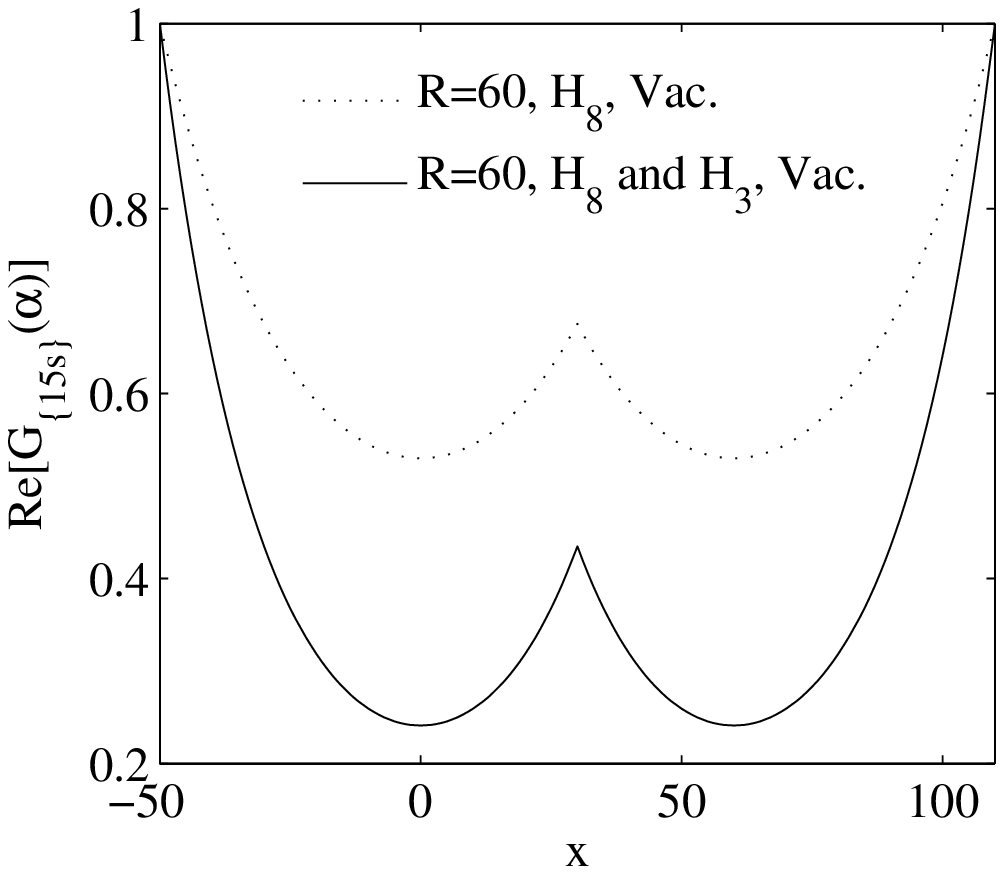}
\caption{ a) The group factors ${\mathrm {Re}}G_{\{15s\}}(\alpha)$ of the center vortices versus $x$ at $R=60$ close to the concavity regime corresponding to Fig. \ref{fig:4}. There are fluctuations with equal large sizes of maxima and minima around any time-like leg within the center vortex group factor corresponding to $H_8$ leading to the concavity behavior in the potentials. Adding $H_3$ contribution to the center vortex group factor, it changes smoothly around any time-like leg removing the concavity in the potential. b) The same as a) but for the vacuum domains where the group factor using only $H_8$ or both Cartan generators changes smoothly. The free parameters are $L_{d}=100$ and $L^{2}_{d}/(2\mu)=4$. }
\label{fig:55}
\end{figure}

Figure \ref{fig:55} a) depicts the group factors induced by center vortices for different Cartan generator contributions in the medium size Wilson loop close to concavity regime (about $R=60$) for the representation $\{15s\}$.  Figure \ref{fig:55} b) is the same as Fig. \ref{fig:55} a) but for the vacuum domains. The Wilson loop legs are located at $x=0$ and $x=60$. For the group factor obtained by center vortices corresponding to only $H_8$, the fluctuations of the center vortex group factor with equal large sizes of maxima and minima around any time-like leg lead to the concavity behavior in the potential while these fluctuations are eliminated somewhat by adding the $H_3$ contribution and the group factor changes almost smoothly around any time-like leg. In addition, as shown in Fig. \ref {fig:55} b) the group factor obtained by vacuum domains using both Cartan generators $H_8$ and $H_3$ changes almost smoothly around any time-like leg. Therefore the concavity of the potential could be removed by including the $H_3$ contribution to the potential. 

\begin{figure}[h!] 
\centering
a)\includegraphics[width=0.48\columnwidth]{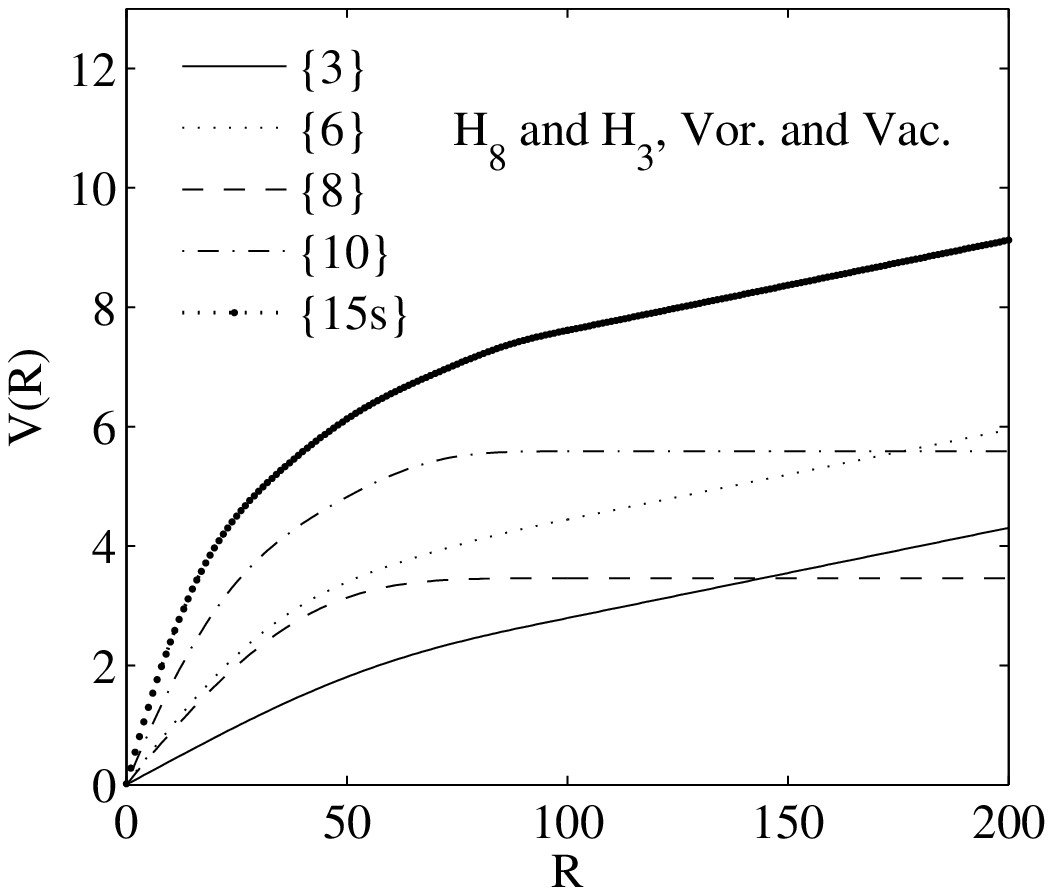}
b)\includegraphics[width=0.48\columnwidth]{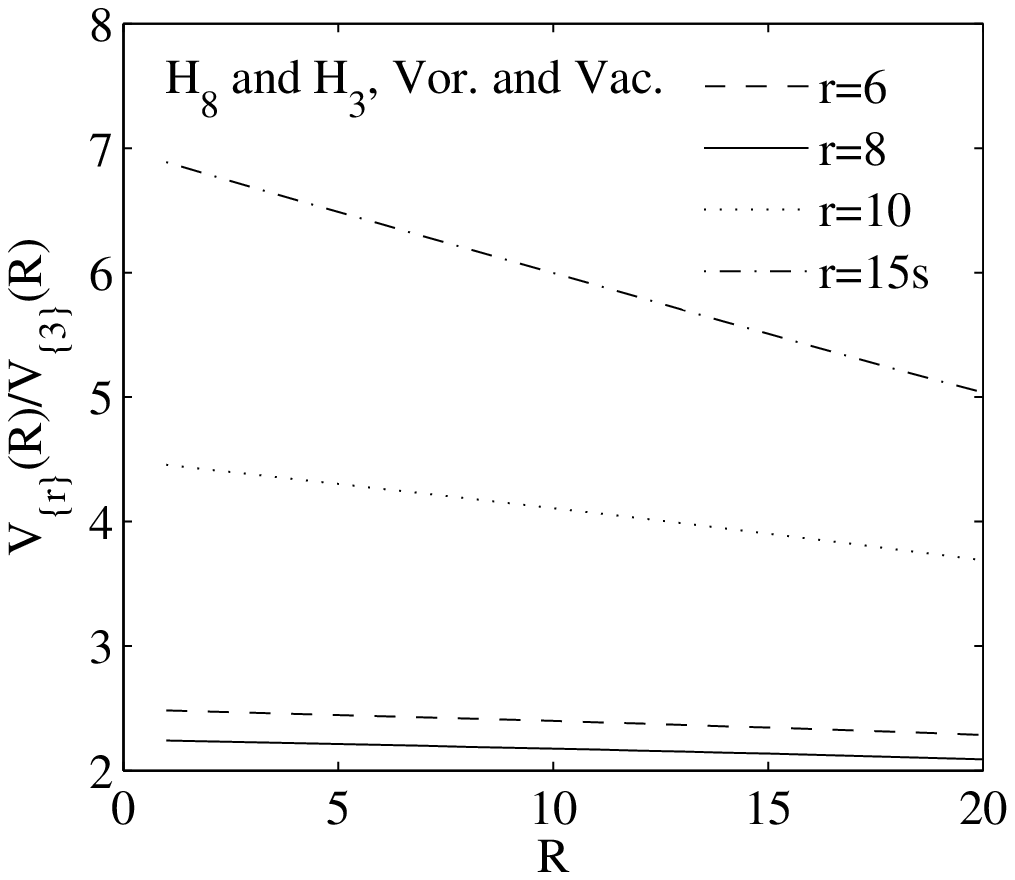}
\caption{ a) The static potentials induced by center vortices and vacuum domains using both Cartan generators for the various representations of $SU(3)$. The satisfactory potentials without any concavity can be achieved for the different representations. b) Potential ratios 
$V_{\{r\}}(R)/V_{\{3\}}(R)$ at the intermediate distances agreeing with the Casimir scaling. The free parameters are $L_{d}=100$, $f_{1}=0.01$, $f_{0}=0.05$, and $L^{2}_{d}/(2\mu)=4$. }
\label{fig:6}
\end{figure}

Figure \ref{fig:6} a) plots the static potentials induced by all domains using both Cartan generators in the various representations for the range $R\in [0,200]$. Furthermore, the potential ratios are shown in Fig. \ref{fig:6} b). Therefore, the satisfactory potentials for
the different representations can be achieved. Using both Cartan generators in the potential is crucial for obtaining different string tensions in the intermediate and large distances in particular for the fundamental representation and also the concavity of the potentials for higher representations can be eliminated. The potential ratios start out at the Casimir ratios ${C_{\{6\}}}/{C_{\{3\}}}=2.5$, ${C_{\{8\}}}/{C_{\{3\}}}=2.25$, ${C_{\{10\}}}/{C_{\{3\}}}=4.5$, and ${C_{\{15s\}}}/{C_{\{3\}}}=7$ and drop slowly from the exact Casimir scaling for all representations. However the potentials induced by the domains corresponding to $H_8$ agree with the Casimir scaling better than those obtained from the domains corresponding to both $H_8$ and $H_3$. 
\section{Conclusion}\label{Sect3}
In the framework of the domain model of center vortices, we analyze the static potentials for various representations in SU($3$) Yang-Mills theory. The contributions of the Cartan generators in potentials between static sources are investigated. Using only one Cartan generator $H_8$ in the potentials, already given as a particular proposal, there are some concavities in the potential for higher representations. In addition, the string tension of the fundamental representation  is the same at Casimir scaling and N-ality regimes while there is no explanation for why the string tension of the fundamental representation should be the same in these regimes. For analyzing these shortcomings, we add the contribution of the other Cartan generator $H_3$ in the potentials and therefore the core of a center domain is corresponding to both Cartan generators. We show that adding $H_3$ contribution to the group factor of the potential the fluctuations of the group factor are removed somewhat and the group factor changes almost smoothly around any time-like leg of the Wilson loop. Therefore, the concavities in the potentials can be eliminated by adding $H_3$ contribution on the potentials. Increasing the number of Cartan generators within the core of domain, the string tension of the fundamental representation at intermediate distances can be achieved different value with the one at asymptotic distances. Although the static potential is governed by only $H_8$ at large regimes, there are the contributions of both Cartan generators at intermediate regimes. Furthermore, we investigate the contributions of the Cartan generators on the Casimir scaling of potentials at intermediate distances. The potentials induced by the domains with the core corresponding to $H_8$ agree with the Casimir scaling better than those obtained from the domains with the core corresponding to both Cartan generators. 

Therefore including all Cartan generators is crucial in obtaining the convex potentials in agreement with Casimir scaling at intermediate regimes and also getting different string tensions at Casimir scaling and N-ality regimes.
\appendix
\section{ Group factors for several representations of SU($3$)}

 The Cartan generators $H_8^{\{r\}}$ and $H_3^{\{r\}}$ for the representation $r$ within the group factors of the static potential given in Eq. (\ref {Vr}) can be obtained using the tensor method. The real part of the group factors using both Cartan generators $H_8^{\{r\}}$ and $H_3^{\{r\}}$ in several representations for a center domain can be calculated as:

\begin{equation}
    {\mathrm {Re}}{G}_{\{3\}}(\alpha^{n})=\frac{1}{3}[cos(\frac{\alpha_3^{n}}{2}+\frac{\alpha_8^{n}}{2\sqrt{3}})+cos(\frac{\alpha_3^{n}}{2}-\frac{\alpha_8^{n}}{2\sqrt{3}})+cos(\frac{\alpha_8^{n}}{\sqrt{3}})],   
\label{group-3}                      
\end{equation}
\begin{equation}
    {\mathrm {Re}}{G}_{\{6\}}(\alpha^{n})=\frac{1}{6}[cos(\alpha_3^{n}+\frac{\alpha_8^{n}}{\sqrt{3}})+cos(\frac{\alpha_8^{n}}{\sqrt{3}})+cos(\frac{\alpha_3^{n}}{2}-\frac{\alpha_8^{n}}{2\sqrt{3}})+cos(\alpha_3^{n}-\frac{\alpha_8^{n}}{\sqrt{3}})+cos(\frac{\alpha_3^{n}}{2}+\frac{\alpha_8^{n}}{2\sqrt{3}})+cos(\frac{2\alpha_8^{n}}{\sqrt{3}})],  
\label{group-6}                       
\end{equation}
\begin{equation}
    {\mathrm {Re}}{G}_{\{8\}}(\alpha^{n})=\frac{1}{8}[2+2cos(\alpha_3^{n})+2cos(\frac{\alpha_3^{n}}{2}+\frac{3\alpha_8^{n}}{2\sqrt{3}})+2cos(\frac{\alpha_3^{n}}{2}-\frac{3\alpha_8^{n}}{2\sqrt{3}})],  
\label{group-8}                       
\end{equation}
\begin{equation}
    {\mathrm {Re}}{G}_{\{10\}}(\alpha^{n})=\frac{1}{10}[1+cos(\frac{3\alpha_3^{n}}{2}+\frac{3\alpha_8^{n}}{2\sqrt{3}})+2cos(\frac{\alpha_3^{n}}{2}+\frac{3\alpha_8^{n}}{2\sqrt{3}})+2cos(\alpha_3^{n})+2cos(\frac{\alpha_3^{n}}{2}-\frac{3\alpha_8^{n}}{2\sqrt{3}})+cos(\frac{3\alpha_3^{n}}{2}-\frac{3\alpha_8^{n}}{2\sqrt{3}})+cos(\frac{3\alpha_8^{n}}{\sqrt{3}})],   
\label{group-10}                      
\end{equation}
\begin{equation}\begin{aligned}
    {\mathrm {Re}}{G}_{\{15s\}}(\alpha^{n})&=\frac{1}{15}[cos(2\alpha_3^{n}+\frac{2\alpha_8^{n}}{\sqrt{3}})+cos(\alpha_3^{n}+\frac{2\alpha_8^{n}}{\sqrt{3}})+cos(\frac{3\alpha_3^{n}}{2}+\frac{\alpha_8^{n}}{2\sqrt{3}})+cos(-\alpha_3^{n}+\frac{2\alpha_8^{n}}{\sqrt{3}})+cos(2\alpha_3^{n}-\frac{2\alpha_8^{n}}{\sqrt{3}})+\\
&+cos(\frac{3\alpha_3^{n}}{2}-\frac{\alpha_8^{n}}{2\sqrt{3}})+cos(\frac{\alpha_3^{n}}{2}-\frac{5\alpha_8^{n}}{2\sqrt{3}})+cos(\frac{\alpha_3^{n}}{2}+\frac{5\alpha_8^{n}}{2\sqrt{3}})+cos(\frac{4\alpha_8^{n}}{\sqrt{3}})+cos(\frac{2\alpha_8^{n}}{\sqrt{3}})+\\
&+cos(\alpha_3^{n}-\frac{\alpha_8^{n}}{\sqrt{3}})+cos(\alpha_3^{n}+\frac{\alpha_8^{n}}{\sqrt{3}})+cos(\frac{\alpha_3^{n}}{2}+\frac{\alpha_8^{n}}{2\sqrt{3}})+cos(\frac{\alpha_3^{n}}{2}-\frac{\alpha_8^{n}}{2\sqrt{3}})+cos(\frac{\alpha_8^{n}}{\sqrt{3}})].   
\label{group-15}                      
\end{aligned}\end{equation}

 %----------------------------------------
% References
%----------------------------------------

\end{document}